**Journal of Financial Risk Management, 2025, 14(4), 428-455**
https://www.scirp.org/journal/jfrm
ISSN Online: 2167-9541
ISSN Print: 2167-9533

![Scientific Research Publishing]

# Governance, Risk, and Regulation: A Framework for Improving Efficiency in Kenyan Pension Funds

## Sylvester Willys Namagwa

Department of Finance and Accounting, Faculty of Business and Management Science, University of Nairobi, Nairobi, Kenya
Email: sylvesternamagwa@gmail.com

**How to cite this paper:** Namagwa, S. W. (2025). Governance, Risk, and Regulation: A Framework for Improving Efficiency in Kenyan Pension Funds. *Journal of Financial Risk Management, 14,* 428-455.
https://doi.org/10.4236/jfrm.2025.144023

**Received:** September 29, 2025
**Accepted:** November 11, 2025
**Published:** November 14, 2025

## Abstract

As life expectancy in Kenya increases, so does the need for efficient pension schemes that can secure a dignified retirement and protect members from old-age poverty. Limited research, however, has explored the efficiency of these schemes under existing governance structures. This study addresses that gap by examining the combined effects of corporate governance, risk management, and industry regulation on pension scheme efficiency in Kenya. Using a quantitative design, we conducted a panel regression analysis on a seven-year secondary dataset of 128 Kenyan pension schemes, totalling 896 observations. Our results reveal significant insights: the presence of employee representatives on the board and effective risk management have a significant positive effect on efficiency. Conversely, independent board members exhibit a significant negative effect. Other factors, including top management representation, female board members, and industry regulation, showed no significant effect on efficiency in the joint model. These findings suggest that the impact of governance and risk management on efficiency is nuanced, with specific factors—like employee representation—playing a more prominent role. We propose that the electoral process for employee board members may introduce a "self-cleaning mechanism" that progressively enhances scheme efficiency. This mechanism offers a novel theoretical extension of Agency Theory, explaining the convergence of interests between elected trustees and scheme members.

# 1. Introduction

## 1.1. Background

Pension systems play a vital role in not only generating income to protect senior citizens against the indignity of poverty, but also ensuring a steady flow of long-term capital for economic growth. Pension schemes' governance should prioritise the adequacy of pension payouts (Hartley & Abels, 2025) to strengthen social safety nets. The increasing life expectancy in Kenya (World Bank Group, 2019) calls for sustainable measures to cushion the elderly, most of whom depend on pension as their only source of income (Ouma et al., 2025). The clamour for contributory pension plans, owing to their sustainability and intergenerational fairness (OECD, 2023), has fuelled the proliferation of Defined Contribution (DC) plans against the Defined Benefit (DB) plans. However, this shift evokes governance concerns, as it burdens the employees (scheme members) with the role of ensuring that their schemes perform well to generate adequate retirement income (Jefferson, 2023). Unlike the DB plan where the employer's obligation endures to the retirement date of the employee, in the DC, the employer's obligation ends once they remit their portion into the pension scheme managed by an independent Board of Trustees.

Boards of trustees, as structured along Corporate governance (CG) best practice as the tool by which providers of inputs for the firm make its managers accountable and responsive to their expectations (Namagwa et al., 2024a), should help scheme members gain the best returns on their savings. Trustees, as agents, have a fiduciary duty to govern schemes in the best interest of their members by formulating the right policies, managing finances, appointing capable service providers, etc. Hence, the size of members' wealth basket depends mainly on how well Trustees invest assets of the scheme (Jefferson, 2023), comply with regulators to avoid penalties, manage risks to exploit opportunities (Katto & Musaali, 2019), and ultimately, minimise losses. CG catalyses the interplay of mixt factors that define the effectiveness of the board in designing oversight tools and their costs to promote trust, accountability, transparency, and mitigate conflicts as envisaged in agency theory (Jensen & Meckling, 1976). Globally, CG is evolving from limiting itself to firm owners and managers, and extending its reach to a matrix of stakeholders in the firm's catchment area, besides breeding corporate integrity as its emerging offshoot for ethical practices (Vu et al., 2025).

Regulation sets compliance requirements for stakeholders in an industry to safeguard their interests and protect it from the dangers of market imperfections, moral hazard, information asymmetry, and inefficiency (Namagwa et al., 2024b). Regulation is devolving from government agencies to industry players under the aegis of self-regulatory bodies (SRBs) since stringent government interventions inhibits efficiency (Jefferson, 2023). However, for the risk of regulatory capture, SRBs are discouraged in complex and high-risk areas like regulation of virtual assets against money laundering (Schwarz & Chen, 2021). Sustainability factors through environmental, social, and governance concerns have been coopted into regulation to influence risk-based investment strategies that cover the climate change risk (Bocchialini et al., 2025; Katto & Musaali, 2019). Risk management include the initiatives that firms take to price, influence, and reward their exposure to risks, creating a reasonable assurance that goals of the firm would be achieved affordably. Pension schemes court more risks due to their long-term investment horizons (Jefferson, 2023) and they therefore need to derive an optimal trade-off between the cost of risk control and the likely losses from risk events. They thus integrate sustainability into their initiatives for resilience and better long-term performance (Anton et al., 2025).

Pension schemes should constantly pursue efficiency to maximise returns for their members. Efficiency is an optimization solution consisting of sustainable firm activities that build the highest value of human, economic, social, and environmental benefits from a given set of inputs. A firm achieves efficiency when it keeps its costs at the lowest level possible while maximising its returns sustainably. Efficiency guides investors in appraising opportunities across firms so that the most efficient firms attract better prospects (Bocchialini et al., 2025). In their quest to retain pre-retirement lifestyles, members of pension schemes pursue efficient scheme operations to accrue the highest possible return on their contributions. Globally, efficiency is being adopted as a key predictor of productive performance as it carries a realistic approach to measuring performance and its ratios capture the effect of unproductive decisions that singular financial ratios ignore (Shabbir et al., 2020). It also illuminates long-term sustainability of firms (Kariuki, 2023) to guide the actuarial fairness across the demographic spread of the scheme membership.

## 1.2. Problem Statement

The global financial crisis at the outset of this century exposed significant weaknesses in CG systems that resulted into widespread corporate failures and substantial investor losses. Kenya has had its share of corporate failures across various sectors, resulting in significant financial losses for investors and hefty penalties for culprits (Kariuki, 2023). Governance lapses in pension schemes not only expose scheme members to old age hardship, but also dampen the public trust in the pension systems. The growing complexities in the structure, products, and regulatory regimes of the modern corporation threatens the traditional approach to CG that focuses on the pact between owners and managers (Shabbir et al., 2020). For instance, Trustees in pension schemes hire expert service providers who shape the risk profile of their schemes but with no direct accountabilities to scheme members. Further, this expert service comes at a cost on members' wealth without a guaranteed equivalent return (Kimeli & Wepukhulu, 2018). Moreover, some scheme managers make suboptimal decisions out of conflict of interest (Waweru, 2021) and chip away scheme members' wealth without any consequences.

Globally, studies on CG, risk management, regulation, and efficiency of firms arrive at inconclusive results. The reasons include the approaches used to opera-tionalise the variables, the choice of variables, the econometric models adopted, and the divergences that obtain at contextualization of the studies. Furthermore, many of these studies have been undertaken in European and Asian countries – jurisdictions that exhibit contextual differences with Kenya in terms of technological sophistry, social security culture, maturity of industries, regulatory environment, and the level of economic development. These variations give rise to contextual gaps that make findings in those jurisdictions not directly fit into the Kenyan setting. For example, Papik & Papikova (2021) find a negative relationship between regulation and performance of pension funds in Slovakia. However, the Slovak pension funds are closed, such that once an individual joins, they cannot exit, unlike the Kenyan case where members of pension schemes leave when they lose their jobs or can transfer their savings to other schemes on notice.

Papik &Papikova (2021) further exhibit a conceptual gap in the focus on regulation and firm performance without considering the effect of other variables in value-web that would play a crucial role in that relationship. The other gap is methodological, firstly in the use of weak performance models, and secondly in the use of a small sample in a heterogeneous population of funds that operate under different regulators of the exclusive investment vehicles. Kowalewski (2012) finds a positive link between performance of pension funds and board composition in Poland. Yet, this study creates a conceptual gap by omitting risk management as variable, given its critical status as a board-level commitment in defining a firm's risk management strategy. The contextual gap emerges on the disparity in market sophistication and culture between Kenya and Poland. Shabbir et al. (2020) find a positive link between CG and efficiency in China, yet the study does not examine the role of industry regulation, despite China Securities Regulation Commission and other State agencies being deeply involved in the regulation of all firms operating in China (Tricker, 2021) the effect of which should not be ignored.

Kenyan studies show no unanimity either as of the kind of the relationship existing among these variables. For example, Songa et al. (2025) find that macro-environment significantly moderates the link between CG and performance of commercial State firms. This study, however, presents a methodological gap given its use of cross-sectional data, which does not reveal dynamic interactions of variables across periods. Ondiba et al. (2024) find that risk management has a significant positive influence on performance of agricultural State firms. Still, the study reflects contextual dissimilarity with the pension industry in operations and the regulatory framework. Kiptoo et al. (2021) find a positive link between CG and performance; however, the study is limited to insurance firms just like the case of Kariuki (2023) who finds mixed results when examining the effect of different aspects of CG on efficiency of insurance firms. What, then, is the joint effect among CG, risk management, and industry regulation on the efficiency of pension schemes in Kenya? The remainder of this article will first present a conceptual framework, a comprehensive review of existing empirical literature on the topic, then detail the research methodology used, followed by a presentation of the findings and their discussion, concluding with recommendations based on the results and identifying areas for further investigation.

## 2. Conceptual Framework

### 2.1. Corporate Governance

CG is an instrument of power over firms, and it includes what the board does and how it affects stakeholders even though its focal point is the relationship between the firm and its members. It enlists globally integrated best practices to help the board comply with the firm's regulators and assure its members of robust risk management devices that protect firm owners' interests through apportioning accountabilities across stakeholders (Masanja, 2021). Poor CG practices exposes a firm to inefficiency through managerial shirking, suboptimal decisions, and even outright malfeasance that hurts stakeholders' interests (Sharma et al., 2021) and ultimately invites market response (Vu et al., 2025). The structure of CG in Kenyan pension schemes is different from other firms. Firstly, the regulator determines the approach of their governance systems as a way of protecting members' wealth from abuse (Ouma et al., 2025). Secondly, pension schemes run under Boards of Trustees based on Trust arrangements in which Trustees hold members' assets as fiduciaries with a higher standard of care is bestowed upon them. A Trustee's term can be renewed by either re-election by the members or re-appointment by the scheme sponsor (Losialoi et al., 2024); re-elected Trustees invariably demonstrate their performance record to their constituents to secure a subsequent term. Thirdly, regulators require Trustees to meet suitability criteria for appointment, and finally, the concept of ownership of scheme assets is spread across its current and future members, including heirs through succession (Bonyi & Stewart, 2019).

The Trust arrangement makes CG a powerful driver of efficiency in pension schemes. Boards play a pivotal role in determining the investment trajectory of their schemes (Bocchialini et al., 2025) as they define which matters can be delegated for which they appoint professional service providers albeit under their monitoring and appraisal. They also ensure that members receive regular updates on the scheme's performance besides disbursing benefits when they fall due (Losialoi et al., 2024). Studies operationalise CG on board roles and features. Kariuki (2023) defines it on board independence, gender diversity, audit quality, board size, CEO duality, and intensity of board activities. Kaur &Vij (2017) use board independence and size, CEO reward, and ownership structure. Sharma et al. (2021) use a weighted average score on transparency, disclosure, board responsibilities, the role of stakeholders, rights of shareholders and their equitable treatment. Khan et al. (2018) use an index of board shareholdings, transparency, and disclosure. Sakawa & Watanabel (2020) use the monitoring role of diverse shareholders. Katto & Musaali (2019) suggest definition upon the board's decision-making process, competency, and its avenues of accountability to members.

The fiduciary role of Trustees binds the Board to act legally and ethically in the best interest of the scheme members and their beneficiaries, by ensuring the long-term financial security of their retirement savings. This study, therefore, operationalises CG through a set of board characteristics. These are the proportions of the board in terms of top management, employee representatives (scheme members elected to the board of Trustees by their colleagues), female board members, and independent board members. The study conceived that operationalising CG through varying proportions of the board characteristics should create different governance scenarios, reflecting subtle inputs of Trustees in mitigating agency problems from different perspectives for a deeper insight.

## 2.2. Risk Management

Risk management is a key deliverable for the corporate board. The board, through corporate governance, catalyses sound business decisions that essentially promote efficiency of firms by creating a platform upon which a firm's risk management framework stands. Through risk management, a firm is able to communicate across its functional areas the level of risks acceptable and the specific modalities of defining and measuring them and coming up with their management plans. Variability of returns threatens the actuarial plans of pension schemes (Ouma et al., 2025) and the persistence of this variability might upset the actuarial fairness of schemes to their membership or beneficiaries. Trustees assess and approve the investment policies and strategies for the scheme's assets, ensuring they align with the scheme's objectives and risk tolerance. Consequently, Valaskova et al. (2018) contend that risk management is a source of information for making rational decisions that target higher levels of efficiency.

Risk management is key in this study because pension schemes are chiefly involved in long-term investment ventures that come with a possibility of asset values varying from their projections. Sticking to good governance practices promotes the effectiveness of the scheme's risk management framework, especially with the exposure of schemes to liquidity risks when there is a mismatch between asset maturities and the crystallization of retirement benefits, including those caused by short-notice withdrawals. When Trustees effectively manage risks, positive returns accrue for members and the economy in general (Katto & Musaali, 2019; Ouma et al., 2025). However, Kiwanuka (2019) cautions that the effectiveness of the Trustees to manage risks depends majorly on their skillsets and the regulation regime under which they operate. Notably, risk management is sector sensitive (Anton et al., 2025) and thus should be operationalised along various typologies depending on the unique circumstances of a given business or industry.

Florio & Leoni (2017) operationalise risk management through the presence of Chief Risk Officer (CRO), risk committee of the board and its frequency of reporting to the whole board. Masanja (2021) operationalizes risk management upon the existence of risk management infrastructure and management oversight culture supported by segregation of tasks and control monitoring systems. Gon-zález et al. (2020) argue that risk management should be operationalised through the existence of CRO, risk committee, and insurance or sponsor's guarantee against financial loss. This study operationalise risk management along the thoughts of Masanja (2021), on the development of risk management infrastructure i.e. the management oversight & culture, strategy & risk assessment, control systems, information, reporting and communication. These parameters are summarised into two: risk oversight structure and risk reporting system – this is because risk management requires infrastructure for constant oversight and a clear reporting system that makes it easy to determine the level of implementation of a scheme's risk management system and evaluate its quality. Anton et al. (2025) observe that an ERM framework should have elements that demonstrate resource commitment to its success.

### 2.3. Industry Regulation

Industry regulation consists of efforts by a government agency or other organized institutions to protect interests of stakeholders of an industry through sustaining its growth and stability. It may also involve setting laws, guidelines, rules, standards, or practices with which all participants in the industry must conform (Bonyi & Stewart, 2019). In the pension industry, regulation involves the act of the regulator in monitoring industry players' investment choices, risk management practices and adherence to set guidelines (Ouma et al., 2025). It creates mechanisms to level out market imperfections, discourage moral hazard, eliminate consumer myopia, and encourage fair competition. In this way, it reinforces corporate risk management such that schemes enjoying good CG would prioritize risk mitigation to comply with regulatory requirements. It ensures stability, order, and predictability of the business thus building confidence among stakeholders to entrust their assets to the industry (OECD, 2025). The regulator registers and licenses key industry players to ensure they meet the provisions of the law, and monitors their operations to protect the interests of scheme members and sponsors.

Trustees, in the course of their fiduciary duties, carefully select consultants based on the scheme's risk management framework and in compliance with the requirements of the regulator, thus, avoiding the costs of non-compliance. This interplay demonstrates the moderating and intervening effects of risk management and industry regulation to the link between CG and efficiency of pension schemes. The trend of CG among pension schemes determines the approach taken by the regulator to supervise them. This means that well governed schemes would have fewer compliance inspections from the regulator and therefore concentrate on value adding business activities (Katto & Musaali, 2019). Industry regulation streamlines the administration of pension schemes through the payment of benefits, asset valuation, financial reporting, and investment guidelines. It thus provides some flexibility in allowing pension schemes to set their own governance rules in a Trust Deed with a caveat that they do not prejudice the interests of their members. In this manner, they reinforce CG by ensuring that schemes commit to universally recognized behaviour while at the same time allowing stakeholders to make meaningful scrutiny of the schemes' actions, their economic fundamentals, and any other information useful to their business. Regulation thus influences investment decisions of firms (Bocchialini et al., 2025), which at times work against the innovative interests of the firm.

An effective regulation model should anticipate turbulence that comes with financial crises or disruptive technology as a form of risk management. Indeed, Tricker (2021) holds that rigid regulation regimes deny firms the flexibility to accommodate the changing market environment which can easily result into opportunity costs. Similarly, regulation comes with a cost-benefit trade-off that requires regulators to shape effective regulatory frameworks that allow space for markets to function effectively and to respond to emerging expectations of stakeholders (OECD, 2025). This flexibility should come with transparency as a key regulatory component that helps in bolstering risk management regimes and efficiency, as operations of the firm are opened up for scrutiny and eventual market response (Anton et al., 2025). Operationalizing industry regulation may take a variety of approaches. Drahos (2017) vouch for decision-making process, behaviour of investment managers, and the quantitative portfolio restrictions. Akande & McMillan (2018) use liquidity, capital, and asset quality restrictions. Papik & Papikova (2021) use contribution levels and exit options. The current study use the scheme's compliance level to the regulator's conditions proxied by policy documents maintained as required by the regulator.

## 2.4. Firm Efficiency

Firm efficiency is characterised by the effective use of firm resources to create more revenues – stretching beyond economic returns to cover all the value generated for stakeholders. Efficiency focuses on optimizing resource utilization to produce the maximum output with the least input, which is a key driver of performance. Firm efficiency is an important component of firm performance as it demonstrates the marginal performance output a firm can extract in a set of inputs (Vu et al., 2025). CG practices can significantly enhance efficiency of pension schemes when boards champion transparency, accountability, sound decision-making, and risk management initiatives to reduce agency costs (Waweru, 2021). However, the effectiveness of the CG mechanism depends on the diversity and competence of the board, the robustness of ERM frameworks, and the enduring compliance with regulations. Strong CG regime fosters stakeholder confidence and attracts investments, ultimately contributing to efficiency and the long-term sustainability of the schemes.

The current developments in firm efficiency include enhanced regulatory oversight, progress in investment strategies, and the adoption of technology to improve operational efficiency and transparency (Vu et al., 2025). In the pension industry, efficiency is increasingly being recognized as a critical measure of performance, with a growing focus on how effectively pension schemes convert con-tributions and investments into retirement benefits. Efficiency drives pension schemes to achieve better returns for their members while ensuring actuarial fairness across the demographic spread of the scheme membership. Efficiency can be measured through Data Envelopment Analysis (DEA) or Stochastic Frontier Analysis (SFA). DEA is a non-parametric method used to evaluate the relative efficiency of decision-making units (DMUs) by comparing them to a "best-practice" frontier. Recent advancements in DEA include the use of super-efficiency approaches and the integration of environmental considerations into the analysis (Kariuki, 2023). On the other hand, SFA is a parametric technique that estimates efficiency by separating the effects of inefficiency from random error. Recent developments in SFA include the use of random coefficients to account for firm heterogeneity (Vu et al., 2025).

## 2.5. Pension Schemes in Kenya

The pension industry in Kenya has about 4 million members out of the 20 million working population and is worth USD 17 billion, about 15% of the country's GDP (KNBS, 2025). This industry is regulated by the Retirement Benefits Authority (RBA) under the functional supervisory model and is one of the largest pools of investment assets in the economy with over 50% of its assets invested in government securities, 19% in guaranteed funds, 11% in immovable property and 9% in quoted equities. The past few years have seen pension schemes diversifying their investments into alternative investment classes which include offshore investments, unquoted equities, private equity, real estate investment trusts (REITs), commercial paper, and non-listed bonds. Among the alternative investments classes pursued in the recent past, all except REITs registered remarkable growth averaging 80% (RBA, 2024).

The pension industry in Kenya dates back to 1921 upon the founding of a pension plan for European public officers working in Kenya. When Africans started joining formal employment, another pension plan was established for them in 1946. The two plans drew funds directly from the Consolidated Fund. The earliest contributory pension plans were made upon the establishment of the National Social Security Fund in 1965, as a provident scheme for all workers in Kenya, and the Widows and Children Pension Scheme in 1966 to cover families of male government officers only. More contributory pension schemes started emerging when senior officers in the private sector found that the contribution rates at the NSSF were too low to provide them with sufficient income replacement upon retirement (National Treasury, 2023).

Pension schemes in Kenya are classified in various ways: on product type, there are Individual, Umbrella, or Occupational schemes. Individual schemes enlist members at their personal level, Umbrella schemes draw members from smaller schemes under various employers, while Occupational schemes draw their membership from one employer. Schemes are classified under funding type as contributory or non-contributory depending on who, between the employer and the em-ployee, takes the funding responsibility. Basing on their funding structure, schemes are either DC or DB. In a DC, a member contributes to the fund in proportion of their pensionable earnings, the scheme then invests the contributions to earn a return against which retirement benefits are availed net of the expenses incurred to run the scheme and or manage the asset to maturity. In a DB, the employer, based on predefined rules, finances it and the ultimate payout is actuarially determined based on one's final salary and years of service (Ouma et al., 2025).

Schemes are also be classified along investment type as guaranteed, segregated, or hybrid schemes. Guaranteed schemes assure members the value of interest and capital while segregated schemes run a designated market-indexed fund structuring its investment profile along the scheme membership. Hybrid schemes mix both approaches. At a basic level, NSSF remains the mandatory national pension scheme for all Kenyan workers, into which members and their employers contribute 12% of an employee's pensionable earnings in equal proportions. Workers with alternative pension schemes recognized by the RBA can opt out of the NSSF's second tier contributions but retain their membership for the first tier if their contributions exceed the gazetted rates, among other requirements (Losialoi et al., 2024).

## 3. Literature Review

### 3.1. Theoretical Review

The corporate board needs a robust theoretical framework to buttress its decision-making as summarised in this section. Agency theory provides a platform upon which the board justifies its actions of increasing the value of the firm. Financial distress theory provides a risk management approach to guard a firm's solvency, while stewardship theory holds that managers prioritise the interests of owners, and stakeholder theory takes a societal perspective of social responsibility.

#### 3.1.1. Agency Theory

Smith (1776) questioned how directors would pursue interests of the firm owners just as they would do to their own without a bridle to restrain their selfish instincts. The theory envisages a relationship where the principal appoints an agent to work and make decisions on their behalf. This arrangement breeds the agency dilemma, which runs on the idea that agents as utility maximizers seek to get the best for themselves and therefore cannot win the trust of their principals (Jensen & Meckling, 1976). It serves firms in which ownership and management are autonomous and survive on the strength arising from well-crafted contracts that allow segregation of duties and promotion of honesty, transparency and accountability (Tricker, 2021), however, these contracts come at a cost. The governance of pension schemes in Kenya exceptionally runs against the background of multiple specialist consultants—actuaries, fund managers, custodians, approved issuers, administrators, advisors, and auditors (Losialoi et al., 2024)—all of whom deter-mine the efficiency and the risk profile of the client scheme.

The inclusion of multiple consultants as a key regulatory aspect of the schemes compounds the agency problem and breeds a variant of agency problem where the owners (members) appoint agents (the scheme) to administer their wealth, but the scheme in turn appoints other agents (service providers/consultants) to tend this wealth on their behalf in a contractual arrangement. This creates multiple layers of agency dilemma, giving risk management a multidimensional approach across the many participants along the pension business value chain (Bocchialini et al., 2025), stifling information flow, typified by the agency theory, and ultimately determining the quality of decisions in terms of their attendant payoffs. The theory anchors this study as it offers a framework of resolving conflicts emerging out of the agency dilemmas among the many nodes of decision making and settling the costs of managing such conflicts. Indeed, there are several agreements between Trustees and other parties, which draw from the perspectives of this theory, for instance, the agreement for provision of administration services, the agreement with fund managers for investment of scheme funds, the agreement with custodians of scheme assets and documents, among other agreements for segregated roles in the schemes (Losialoi et al., 2024).

Therefore, agency theory runs a bipartite model that disenfranchises stakeholders outside the agent-principal axis and essentially worsens the separation of decision-making from risk bearing as manifested in the service agreements mentioned, where decision makers do not bear the risk of their decisions (Tricker, 2021) yet their decisions structure risk profiles and efficiency of the client schemes. Some professional service providers in Kenya have been accused of trading scheme assets at a discount and repurchasing them at a premium shortly thereafter, without the knowledge of scheme members. Such actions result into moral hazard in not only pension schemes and but also in highly leveraged firms like banks, where creditors hold a bigger stake, but are not involved in such decisions. Further, agency theory's bipartite model flaws its application in the governance of firms with complex ownership structure owing to its assumption that shareholders have homogeneous interests, which turns out to be untrue given the variety of shareholders ranging from activists to venture capitalists who join firms with diverse objectives. It further ignores board behaviour, which plays an important role in the adoption of corporate risk management strategy (Pande & Ansari, 2014).

Concisely, agency theory postulates conflicts of interest among the many nodes within the governance framework of pension schemes and is concerned with managerial discretion and opportunism. The conflicts among consultants and Trustees are behavioural and affect the cost structure of the schemes and ultimately their efficiency. Trustees, for example, can exploit their fiduciary position to enjoy unsanctioned benefits at the expense of the scheme members. Similarly, the contracted fund managers can invest scheme members' funds in suboptimal projects or in extremely risky ventures while pursuing bonuses or dishonest ends. Fund administrators, on the other hand, can fabricate transactions to inflate management fees drawn out of scheme members' wealth. These activities by agents have a cost element, which affects the overall efficiency of these schemes and the wealth of the principals—scheme members. Within a pension scheme's governance structure, members exercise control over their elected trustees through periodic elections, a mechanism that acts as an extension of agency theory. This process incentivizes trustees to perform effectively and provides an internal control mechanism for accountability: should they fail to provide adequate oversight, mismanage strategy, or cause a decline in member assets, the members have the power to replace them. This electoral power creates a self-cleaning effect on the board composition.

### 3.1.2. Transaction Cost Theory

This theory holds that the optimum structure of the firm is one that minimizes the cost of transacting such that firms can derive economies of scale from the governance structures they adopt (Coase, 1937; Williamson,1979). That a firm saves costs if it runs its activities from within itself. However, as it grows in size, it becomes cheaper to outsource. A firm refines its cost of checks and balances around audits, disclosure of information, segregation of duties, risk analysis, and the structure of the board (Pande & Ansari, 2014). The firm therefore spends on these structures to a level where the increase in costs matches the reduction of the possible loss from non-compliance. This theory may fail where other factors dictate a firm's governance structure whose adjustment cannot possibly happen to accommodate efficiency needs. For example, to comply with regulatory requirements, pension schemes are structured to have many players along their value web—trustees, regulators, fund managers, fund administrators, actuaries, legal advisers, auditors and custodians (Losialoi et al., 2024). They therefore suffer asset specificity because they must include these parties in their operations, which gives rise to costs whose control, ironically, is incumbent upon the governance structure of the scheme itself. Further, there is no clarity on who bears the greatest responsibility and what is the ideal structure of monitoring the scheme's operational strategy (Katto & Musaali, 2019).

This theory flounders also, where decisions are affected by opportunism and bounded rationality, where stakeholders choose to pursue selfish goals against the best interests of the firm, or they lack sufficient information or skills necessary to harness the desired cost outcomes. The regulatory system of pension schemes in Kenya allows members to nominate their colleagues to sit on the Board of Trustees. However, some of these representatives could come in without proper skillset for better cost decisions advocated by this theory. Finally, the advent of complex trading platforms like hedge funds or derivative markets breeds the intricacy of identifying the owners' best interest and the agent's accountabilities promoted here. Nevertheless, transaction cost theory pursues the most suitable governance structure for the best outcomes. In pension schemes, lower costs means higher efficiency and more wealth for members. This theory, thus, vindicates the emer-gence of umbrella schemes, which are built to ride on the economies of scale, besides the cost allocation decisions made among the multiple parties in the schemes' value web in quest for efficiency.

### 3.1.3. Stewardship Theory

The theory postulates that there should be no dichotomy of roles between owners of the firm and their managers since no conflict exists between them (Donaldson & Davis, 1991). It gives the legal foundation for corporate entities that managers have a duty to the owners and not unto themselves or other interest groups (Tricker, 2021). This theory is defined by a variant of altruism where managers' motives are not personal, but follow the goals and aspirations of their principals, an action that affords them satisfaction (Pande & Ansari, 2014). The theory advocates for governance structures that promote effective coordination, with the reasoning that, adopting facilitative structures lead to efficiency (Davis et al., 1997). Due to complexities in defining ownership, shareholders in the contemporary corporation have become too remote to be involved in the selection of directors as this theory avers. Pension schemes' substructure compromises the accountability of Trustees, such that when their interests do not align with those of scheme members, they invariably use the scheme's resources to shield themselves from accountability (Tricker, 2021). Intricacies around financial firms and their system of regulation may not favour selflessness that this theory espouses, mainly owing to the pole position that segregation of duties serves in internal controls (Sharma et al., 2021). Its collaborative approach can prove inefficient since all the decisions in the firm would involve convergence of all parties (Keay, 2017).

### 3.1.4. Stakeholder Theory

The theory presents a model describing a firm as a collection of both competitive and collaborative interests with fundamental value (Donaldson & Preston, 1995) where a firm invariably seeks concessions within its environment to ensure that each of its interested parties receive a certain degree of satisfaction even without a legal structure to endorse their relationship. Stakeholders are essential to the success of the firm (Sakawa & Watanabel, 2020) and thus, this theory helps in handling stakeholder tensions to create a conducive business environment and a sustainable economic system upon which the firm derives its own survival through external dependencies. The diverse pursuits of stakeholders complicates the application of this theory since firms would find it difficult to balance multiple and potentially conflicting interests. The theory ignores the necessity to protect corporate players against unsustainable stakeholder engagements that may compromise their profitability, hence, boards may hesitate to prioritise interests of stakeholders that tend to undermine the efficiency of the firm itself.

Moreover, it is difficult to define interests of all stakeholders or even maximise their interests concurrently without causing conflicts within the stakeholder groups themselves (Tricker, 2021), a venture that may politicise the firm's business and even handicap it in a competitive business environment (Jensen, 2002).

Stakeholder theory considers the governance structure of pension schemes by recognising the role of every participant in the value web that leads to the success of the scheme. It recognises the role of risk management in protecting the scheme and satisfying the interests of scheme members/beneficiaries and other parties whose prospects depend on the success of the scheme. This theory also considers the need to comply with industry regulators as key stakeholders, which in turn saves the scheme the cost of penalties for non-compliance.

### 3.1.5. Financial Distress Theory

This theory, proposed by Smith & Stulz (1985), holds that firms facing a possibility of financial distress pursue risk management initiatives that link hedging arrangements and managerial compensation. The theory shapes corporate risk management decisions, particularly for firms with long-term investment horizons. Firms can make informed decisions on managing their risks and maximizing their long-term value if they understand the potential costs of distress and the benefits of hedging. Managerial compensation arrangements correspond with the payoffs that accrue to the firm, such that, the manager only takes risks when the consequent reward is higher for both himself and the firm for the reason that his wealth is a function of the firm's value (Drahos, 2017). This theory, however, flops in its assumption that transaction costs are negligible and that the expected returns on all financial assets are equal (Smith & Stulz, 1985), this may not hold in the case of pension schemes where multiple parties with numerous transactions result into costs which can turn out to be significant proportions of the scheme's resources.

Additionally, the regulator requires schemes to diversify their investments across various classes of assets (Ouma et al., 2025) whose returns cannot possibly be equal. Lastly, the financial distress theory exposes firms to short-termism tendencies that might work against the long-term interest of firms, especially those with longer investment horizons like pension schemes whose prospects can be jeopardised by focusing on short-term transactional wins. Nevertheless, this theory considers the role played by risk management in linking CG and firm efficiency against the background of creating sustainable value while managing risks. Pension schemes invest funds in risky assets that require good governance structures backed with regulatory safeguards to reduce the risk of losing scheme members' wealth for which this theory prescribes hedging and compensation arrangements. Concisely, financial liquidity is a critical component of pension schemes for it ensures the fulfilment of the cash promise.

## 3.2. Empirical Review

Kariuki (2023) investigates the association among governance mechanisms and efficiency of insurance companies in Kenya, estimating efficiency using DEA on data drawn from 2013 to 2020 from 53 insurers. The results suggest that CG through board independence, audit quality, and gender diversity have a significant and positive influence on technical efficiency of insurance companies except for board size, which exhibits a significant negative effect. The study presents a contextual gap given the disparate governance structures between the insurance industry and the pension industry. Unlike in pension schemes, where scheme members have slots on the boards of trustees, policyholders do not have similar board opportunities to drive the strategic direction of insurance firms. This makes the results of this study ineligible to be generalized upon the pension industry.

Using DEA and panel data to investigate the link between CG and firm efficiency on 136 non-financial firms listed in Pakistan Stock Exchange from 2008 to 2017, Khan et al. (2018) concludes that CG mechanisms have a positive effect on firm efficiency. The study brings forth a conceptual gap in its failure to consider other variables with intervening/moderating character on the relationship between CG and firm efficiency. A methodological gap emerges in its exclusion of financial firms and unlisted firms. Considering that firms seeking to be listed must fulfil certain governance and performance prerequisites before admission to the bourse (OECD, 2025), the study might have left out diverse firms with heterogeneous operational and governance systems unavailable to the stock exchange players. The current study considers pension schemes of diverse sizes and forms of operations.

Masanja (2021) studies the level to which the board's audit committee contribute to the effectiveness of ERM in social security funds in Tanzania. Using OLS technique on cross-sectional data with 193 observations collected through self-administered questionnaires, the study finds a significant positive relationship between audit committee characteristics/oversight activities and the effectiveness of the ERM. However, this study suffered methodological gaps that come with cross-sectional data e.g. undetected endogeneity finds no chance to be corrected. The current study employs panel data to bridge this gap. Further, the study presents a conceptual gap by ignoring the conceptual necessity of industry regulation in the ERM and CG interplay. Regulation protects rights of stakeholders in the pension industry thus underscoring its importance in their governance (Bonyi & Stewart, 2019).

Zabri et al. (2016) set out to establish the link between CG practices and firm performance focusing on 100 listed companies in Bursa Malaysia from 2008 to 2012. The study uses descriptive and correlation analysis to test the hypothesis and finds that board size has a significant weak negative relationship with ROA and insignificant with ROE. The exclusive sample of top listed firms presents the possibility of sample bias against smaller or unlisted firms that could not make it to the sampling frame, resulting into a methodological gap. In addition, it comes with a conceptual gap, as the study does not use any moderating and intervening variables in its matrix. These variables play a crucial role in understanding the nuances of interactions between variables, and their absence can lead to incomplete or potentially misleading conclusions due to limited depth and breadth of the findings (Lin et al., 2019).

Sharma et al. (2021) studies 11 banks designated as public limited banks to test the effect of CG on their efficiency. The study uses DEA to determine the value of the dependent variable, and the banks' CG scores to define the value of the independent variable. The study covering a period of five years from 2015 to 2019 finds a positive relationship between CG and firm efficiency in the majority of the firms considered. However, some banks whose high CG scores did not reflect higher efficiency presents a knowledge gap. This could mean that there might possibly be other important variable(s) at play that were not explored during this study.

## 4. Research Methodology

### 4.1. The Purpose, Scope and Method of the Research

This paper sets out to investigate the joint effect of CG, risk management, and industry regulation on the efficiency of pension schemes in Kenya. The study relied on positivism philosophy given the use of existing theories and measurement of variables. It adopts a quantitative research design to describe the status of study variables and determine the extent of relationships among them, and panel research design for time and longitudinal dimensions to track performance changes among pension schemes from 2015 to 2021. The study targets a population of the 1189 pension schemes registered by the RBA as of 31st December 2022. Each of the schemes sampled for the study represents the unit of analysis.

The study uses stratified sampling along Occupational, Individual, and Umbrella schemes to sample 128 out of the 1189 schemes in the RBA's register—all providing 896 observations for the study. Even though pension schemes in Kenya present a relatively homogeneous character, this approach is adopted to ensure that the minor variations in schemes along product type are adequately represented (Memon et al., 2020). While sample size is a controversial question in empirical studies, there is no universal sample size that can be prescribed for all studies (Bartlett, Kotrlik, & Higgins, 2001). Sathyanarayana et al. (2024) cautions that approaches for determining an adequate sample size require a careful consideration based on factors unique to a study. The sample size of this study considered the number of variables in the model, the resources available for the study, and the research approach espoused. The study collected secondary data through reviewing various reports and audited financial statements of the sampled schemes.

### 4.2. Estimation of Pension Scheme's Efficiency

Several studies operationalise efficiency using DEA (Kariuki, 2023; Lin et al., 2019; Sharma et al., 2021; Vu et al., 2025). DEA is a linear programming technique that identifies the most efficient decision-making units by allocating their relative efficiency scores basing on their inputs and outputs. DEA runs on three concepts of efficiency. Firstly, there is technical efficiency, where physical inputs are converted into outputs relative to the best practice. Second is the allocative efficiency where inputs, for a given level of output and set of input prices, are selected to minimise the production cost, while assuming full technical efficiency in the entity. The third concept combines both technical efficiency and allocative efficiency to fash-ion cost efficiency. An entity at the peak of best practice is said to be 100% efficient (Kariuki, 2023; Lin et al., 2019). The greatest advantage of DEA is that it can incorporate multiple inputs and outputs to determine efficiency among decision-making units. The current study use DEA with constant returns to scale with the value of administrative and investment management costs as inputs, and the amount of return on investments (ROI), and change in net assets as the outputs. This is owing to managerial discretion in the structuring of administrative and investment management costs. This study focuses on technical efficiency as shown in Table 1.

**Table 1.** Efficiency analysis.

| Variable | Symbol | Description |
|---|---|---|
| **Input Variables** | | |
| Investment Management Costs | Ic | Annual amounts spent to manage investment assets |
| Administrative costs | Ac | Annual amounts spent to run affairs of the scheme |
| **Output Variables** | | |
| Net assets of the scheme | Na | Annual change in the value of the scheme assets |
| Return on Investments | ROI | Annual amounts gained on the initial value of investments |

### 4.3. Operationalisation of Variables

The study employed the four variables with their indicators and measurements as shown in Table 2.

**Table 2.** Operationalisation of variables.

| Variable | Indicator (s) | Measurement (s) |
|---|---|---|
| Corporate Governance | Top Management on board (Mgt) | Ratio of managers on the scheme's board |
| | Employee representatives on the board (Mb) | Ratio of scheme members on the scheme's board |
| | Female board members (Fm) | Ratio of females on the scheme's board |
| | Independent Members (Im) | Ratio of independent members on the scheme's board |
| Risk Management | Development of risk management infrastructure (Rm) | The level of risk management infrastructure (risk management parameters as disclosed) |
| Industry Regulation | Compliance to regulator conditions (IR) | The level of compliance with industry regulation (number of policy documents maintained as required) |
| Efficiency of Pension Schemes | Administrative Costs (Ac) | Amount of administrative costs for the year |
| | Investment costs (Ic) | Amount of investment costs for the year |
| | Amount of Return on Investments (ROI) | Amount of returns on investments for the year |
| | Change in Net Assets of the Scheme (Na) | Amount of change in net assets of the scheme at year end |
| Composite value | EFi | DEA Value for efficiency scores |

To quantify regulatory compliance for each scheme, a score was assigned based on the number of maintained policy documents required by the regulator (1 - 9).

The list of documents included the code of conduct, election rules, trustee remuneration policy, board evaluation procedures, risk management policy, ICT policy, member's scheme booklet, conflict of interest policy, and procurement procedures. A higher count of these documents led to a higher compliance score for the scheme represented on a ratio scale for regression analysis. These describe compliance parameters touching on member information, procurement procedures, and policy documents kept by the scheme. The study selects these parameters because they reflect the scheme's ability to ward off costs of non-compliance since disruptions and penalties levied for non-compliance eat into the resources of pension schemes and as a result reduce their efficiency scores.

Similarly, the development of risk management infrastructure is tracked on the existence of management oversight & culture, strategy & risk assessment, control systems, information, reporting and communication. Each scheme's risk management level was scored from 1 to 5, based on disclosed information about these parameters, with a higher score reflecting a stronger overall risk management framework represented on a ratio scale for regression analysis. These determine whether the scheme's risk management regime has requisite infrastructure for constant oversight and a clear reporting system that reflects the level of implementation of a scheme's risk management system and its quality, and ultimately the scheme's commitment to risk management through resource deployment and entrenchment of the favourable risk management culture.

## 4.4. Estimating the Joint Effect of CG, Risk Management, and Industry Regulation on the Efficiency of Pension Schemes

The study used multiple regression models to assess the joint effect of CG, risk management, and industry regulation on the efficiency of pension schemes.

$$EF_i = \beta_0 + \beta_1 Mgt + \beta_2 Mb + \beta_3 Fm + \beta_4 Im + \beta_5 Rm + \beta_6 IR + \varepsilon_i \qquad (4.1)$$

$$RRAI = \beta_0 + \beta_1 Mgt + \beta_2 Mb + \beta_3 Fm + \beta_4 Im + \beta_5 Rm + \beta_6 IR + \varepsilon_i \qquad (4.2)$$

**where:**

$EF_i/RRAI$: Pension Schemes/technical efficiency Scores

$\beta_0$: Regression constant or intercept

$\beta_i$: Regression coefficients of variable i

$E_i$: error term

Mgt: Top management on board

Mb: Employee representatives on board

Fm: Female board members

Im: Independent board members

The study employs $R^2$ to assess the dependent variable variation due to the effects of the predictor variable. To assess the model fit, the F-Test was used to test the significance of the model while T-Test was used to evaluate the significance of the beta coefficient of the predictor variable. The null hypothesis ($H_{01}$) was rejected if the *p*-value obtained from the joint effect model (considering all three variables together) was less than 0.05. This indicates that the combination of CG, risk management, and industry regulation significantly affects the efficiency of pension schemes.

# 5. Hypothesis Testing, Findings, and Discussion

## 5.1. Hypothesis Testing

This research aims to examine the joint effect of CG, industry regulation, and risk management on the efficiency of pension schemes in Kenya. This objective integrates the various aspects of CG (top management on board, employee representatives, female board members, and independent board members) along with the roles of industry regulation and risk management to determine their combined influence on efficiency.

$H_{01}$: CG, risk management, and industry regulation have no joint effect on the efficiency of pension schemes in Kenya.

## 5.2. Findings

The results of the joint effect regression analysis are presented in Table 3.

Table 3. Joint effect regression results.

| Fixed-effects (within) regression | | Number of obs | | = | 896 |
|---|---|---|---|---|---|
| Group variable: Scheme | | Number of groups | | = | 128 |
| R-sq: | | Obs per group: | | | |
| within = 0.0953 | | min | | = | 7 |
| between = 0.0114 | | avg | | = | 7 |
| overall = 0.0034 | | max | | = | 7 |
| | | F (6, 762) | | = | 13.38 |
| corr(u_i, Xb) = −0.5624 | | Prob > F | | = | 0 |
| Efficiency | Coef. | Std. Err. | t | P > t | [95% Conf. Interval] |
| LNMgtR | −0.05798 | 0.183931 | −0.32 | 0.753 | −0.41905 | 0.303096 |
| LNMbR | 1.022987 | 0.130493 | 7.84 | 0.000 | 0.766819 | 1.279154 |
| LNFmR | 0.06614 | 0.099452 | 0.67 | 0.506 | −0.12909 | 0.261372 |
| LNImR | −0.57445 | 0.242305 | −2.37 | 0.018 | −1.05011 | −0.09878 |
| LNRM | 0.327731 | 0.146367 | 2.24 | 0.025 | 0.0404 | 0.615061 |
| LNIR | 0.099513 | 0.144859 | 0.69 | 0.492 | −0.18486 | 0.383883 |
| _cons | −1.52709 | 0.366538 | −4.17 | 0.000 | −2.24663 | −0.80754 |

The regression model's adjusted R-squared (within) value is 0.0953, indicating that approximately 9.53% of the variability in the efficiency of pension schemes can be explained by the combined effect of the variables included in the model. The overall F-statistic is 13.38, with a $P$-value of 0.000, suggesting that the model is statistically significant and that the joint effect of these variables on efficiency is substantial.

Top Management Trustees (LNMgtR) has a coefficient of −0.05798 with a t-value of −0.32 and a *P*-value of 0.753, indicating that it does not have a statistically significant effect on efficiency in the presence of the other variables. This result suggests that top management representation on the board, when considered alongside other factors, does not significantly influence the efficiency of pension schemes. Member (employee) Trustees (LNMbR), on the other hand, shows a positive and significant effect on efficiency, with a coefficient of 1.022987, a t-value of 7.84, and a *P*-value of 0.000. This finding highlights the crucial role of employee representation in enhancing the efficiency of pension schemes, even when considering the combined effects of CG, industry regulation, and risk management.

Female Trustees (LNFmR) have a coefficient of 0.06614 with a t-value of 0.67 and a *P*-value of 0.506, indicating that their effect on efficiency is statistically insignificant when analysed jointly with other variables. This result suggests that female board representation does not have a significant effect on the efficiency of pension schemes in this combined model. Independent Trustees (LNImR) exhibit a negative and statistically significant effect on efficiency, with a coefficient of -0.57445, a t-value of −2.37, and a *P*-value of 0.018. This negative relationship indicates that a higher proportion of independent Trustees may be associated with lower efficiency in the presence of other variables, which could be due to potential conflicts or differing perspectives that might impede swift decision-making.

Risk Management (LNRM) has a positive and statistically significant effect on efficiency, with a coefficient of 0.327731, a t-value of 2.24, and a *P*-value of 0.025. This finding underscores the importance of robust risk management practices in improving the efficiency of pension schemes, even when other factors are taken into account. Industry Regulation (LNIR), however, does not show a statistically significant effect on efficiency in the joint model, with a coefficient of 0.099513, a t-value of 0.69, and a *P*-value of 0.492. This suggests that while industry regulation is important, its direct effect on efficiency may be overshadowed by the more immediate effects of CG and risk management practices.

Based on the results presented in <span style="color:red">Table 3</span>, not all aspects of CG, industry regulation, and risk management jointly and significantly influence efficiency; the study rejects the null hypothesis $H_{01}$, which posited that CG, risk management, and industry regulation have no joint effect on the efficiency of pension schemes in Kenya. The findings suggest that while certain factors are crucial, the interplay between these elements is complex and requires careful consideration to optimise efficiency.

### 5.3. Discussion

The study sought to determine how CG, industry regulation, and risk management collectively influence the efficiency of pension schemes in Kenya. The findings reveal that the three factors significantly enhance the efficiency of pension schemes. Notably, employee trustees and risk management practices were identified as key drivers of efficiency when combined with industry regulation. It should be noted that employee trustees serve upon being elected by their colleagues, hence, their tenure is likely determined by the positive value they create for their colleagues. Independent Trustees were found to have a negative effect on efficiency, even in the presence of strong regulatory and risk management frameworks.

The results suggest that independent trustees seem constrained in an environment with risk management and regulatory bridles. This could also be a pointer that some independent trustees, wielding clout within the pension industry, could sit on the board to further sectarian interests against their official brief. Indeed Habtoor et al. (2024) found that independent directors exhibit a negative effect on firm performance when their interests are not aligned with corporate objectives, which is particularly evident in cases where they lack share ownership. Further, independent trustees could be victims of cultural or structural marginalization, which stifles their voices, or even deny them a freehand to perform. For instance, the beneficial effect of board independence on firm performance is often undermined in companies with high levels of family ownership (Singh, 2025).

These findings are consistent with Agency Theory, which suggests that effective governance mechanisms, when supported by regulatory oversight and robust risk management, can align the interests of managers and stakeholders, thereby improving organizational efficiency. In the context of pension schemes, the combined influence of these factors ensures that governance practices are not only sound but also resilient to risks and compliant with regulatory standards. This comprehensive approach to governance mitigates potential agency problems, such as misalignment of interests or opportunistic behaviour, and leads to better outcomes for the schemes' beneficiaries. These findings introduce a component of self-cleaning governance mechanism within the Agency Theory with regard to employee representatives as a form of behaviour control such that, employee representatives risk losing board positions if they don't perform to the satisfaction of their colleagues whose votes determine their tenure.

This self-cleaning governance mechanism could be a superb theoretical offshoot of the Agency Theory to explain the convergence of interests of trustees and their electors—a framework evolved through periodic trustee elections that goad trustees into accountability, efficiency, and integrity to deter corruption, waste, and mismanagement of members' wealth. Consequently, the positive effect of employee representation and the negative influence of independent board members highlight the complexities of aligning interests within these schemes. This finding therefore supports and refines Agency Theory by suggesting that while independent oversight is important, the practical dynamics of board composition require a nuanced approach to truly align management and investor interests through a self-cleaning governance mechanism.

The results of this study align with the findings of and Khan et al. (2018), Kiptoo et al. (2021), and Songa et al. (2025) found a significant positive relationship between CG and performance of commercial State corporations while moderated by macro-environment. Kiptoo et al. (2021) found a positive relationship between governance and performance while Khan et al. (2018) also concluded that good CG positively affects firm efficiency. The current study therefore advances this understanding by incorporating industry regulation and risk management as additional variables that interact with CG to influence efficiency. While Khan et al.'s study focused solely on governance mechanisms whereas Songa et al. considered the macro-environment, the inclusion of regulatory and risk management factors in the current study provides a more holistic view of the factors driving efficiency in pension schemes. This approach addresses the conceptual gap in Khan et al.'s research by demonstrating that governance, regulation, and risk management are interdependent factors that collectively determine organizational efficiency.

The study also relates to Masanja (2021), who found that audit committee characteristics positively influence the effectiveness of Enterprise Risk Management (ERM) in social security funds in Tanzania. Like Masanja's study, the current research highlights the importance of risk management in enhancing governance practices. However, the current study goes further by examining how risk management interacts with CG and regulation to improve efficiency, thereby filling the conceptual gap in Masanja's work, which did not consider the role of regulation. By employing panel data, the current study also addresses the methodological gap in Masanja's research, providing a more dynamic and comprehensive analysis of how these factors interact over time. This study also mirrors Kariuki (2023) who found that CG through board independence, audit quality, and gender diversity have a significant and positive influence on efficiency. Furthermore, the findings contribute to the ongoing discourse on the importance of composite measures of CG, as discussed by Zabri et al. (2016).

While Zabri et al. identified a weak negative relationship between board size and firm performance, the current study emphasises the significance of considering manifold governance factors in tandem with regulatory and risk management frameworks. This broad approach ensures that governance practices are not appraised in isolation but are understood within the wider context of risk management and regulatory dynamics. This approach addresses both the methodological and conceptual gaps in Zabri et al.'s research by providing a more nuanced understanding of how different governance devices, when combined with regulation and risk management, influence organizational efficiency. Finally, the study complements Sharma et al. (2021), who found a positive relationship between CG and firm efficiency in public limited banks, although their study showed some banks with high governance scores not exhibiting higher efficiency. The current study proposes that the existence of other variables, such as regulation and risk management, may explain such discrepancies.

Hence, by integrating these additional factors into the analysis, the study pro-vides a more complete explanation of the variations in efficiency observed among firms with similar governance scores. This finding suggests that high governance scores alone are not sufficient for ensuring efficiency; they must be supported by effective regulation and risk management practices. In conclusion, the findings demonstrate that the combined effect of CG, industry regulation, and risk management is crucial for enhancing the efficiency of pension schemes. The study highlights the importance of a comprehensive approach to governance that integrates regulatory oversight and robust risk management practices. This approach not only aligns with Agency Theory but also addresses key conceptual and methodological gaps in the existing literature by providing a more holistic understanding of the factors that drive efficiency in pension schemes.

## 6. Conclusion and Recommendation

This study sought to examine the collective influence of CG, industry regulation, and risk management on the efficiency of pension schemes in Kenya. The findings indicate that employee trustees (as a component of CG) and risk management practices, in combination with industry regulation, are key drivers of enhanced efficiency. This research vouches for a revised governance approach for Kenyan pension schemes, with the ultimate goal of maximizing returns for members and ensuring their financial security in retirement. The involvement of employee trustees has a positive effect on efficiency, a pointer that their direct interest in the scheme's outcome aligns their decision-making with the best interests of the scheme membership. Conversely, independent trustees (as a component of CG) exhibit a negative effect on efficiency.

While intended to provide objective oversight, independent trustees can be compromised by personal or sectoral interests, or be marginalized by structural barriers. The study confirms that industry regulation is a critical factor for ensuring the efficient and secure operation of pension schemes, therefore, tailored regulatory approaches, which account for the specific characteristics of different types of schemes, are crucial for member protection and scheme sustainability. Risk management emerges as a vital intermediary in the relationship between CG, regulation, and efficiency, thus proper risk management practices enhance decision-making, improve operational efficiency, protect schemes from losses, and bolster stakeholder confidence.

Based on these findings, a raft of recommendations arise for policy makers and scheme managers. Policymakers should reinforce the role of employee trustees, including establishing clear, regulatory-backed guidelines for their election, tenure, and accountability to leverage on the self-cleaning governance mechanism emergent in this study. Further, a regulatory review of the mandate and appointment process for independent trustees is necessary to address their identified negative effect on efficiency. This could involve crafting a Trustee Monitoring Database, a central database to track and collect data on board activities including necessary expertise and performance metrics, to help identify suitable candidates for board appointment, or for delisting those found unsuitable or with misconduct.

Schemes sponsors should consider new engagement protocols for independent trustees, including more rigorous vetting and conducting skills-gap analyses to identify and appoint well aligned trustees. Scheme managers should embed risk management directly into strategic decision-making processes, including investment and member communication strategies. This would involve implementing proactive risk evaluation protocols, such as regular audits and scenario planning, and investment in technology and data analytics to improve the efficiency of risk monitoring and reporting, and to foster a culture of risk awareness. Scheme managers should also optimize trustee engagement by actively empowering them with access to necessary information and training to keep up with the mercurial demands of pension funds' management.

This study underscores the importance of a holistic approach to managing Kenyan pension funds, one that integrates strong CG, effective industry regulation, and robust risk management. By strengthening the role of employee trustees, reforming the engagement of independent trustees, and embedding risk-conscious practices, policymakers and scheme managers can significantly enhance the efficiency of these schemes, ultimately securing a better financial future for their members.

## 7. Areas for Further Research

The pension system's growing complexity, characterized by diverse plans, regulatory frameworks, and investment strategies, necessitates future research. Future studies should explore pension scheme governance at various geographical levels and investigate the impact of other governance variables not included in this study, such as board diversity in terms of age, professional background, or tenure, on the efficiency of pension schemes.

## 8. Limitations of the Study

The low coefficient of determination (R-squared = 9.53%) is a notable limitation. This statistical measure reveals that the independent variables collectively account for only a small percentage of the observed variation in pension scheme efficiency, highlighting that other determinants are at play.

## Conflicts of Interest

The author declares no conflicts of interest regarding the publication of this paper.